\documentstyle[12pt,version2,aps]{revtex}
\begin{document}
\setlength{\baselineskip}{3ex}

\renewcommand{\theequation}{\arabic{section}.\arabic{equation}}
\newcommand{\eqreset}{\setcounter{equation}{0}}
\begin{center}
{\large\bf Universal Finite-Size Scaling Function \\
of
the Ferromagnetic Heisenberg Chain \\
in a Magnetic Field. II \footnote{
{\normalsize Submitted to J.Phys.Soc.Jpn}
}
\\
\vspace{8mm}
----- Nonlinear Susceptibility ----- } \vspace{5mm}\\
 Hiroaki  {\sc Nakamura} ,
 Naomichi {\sc Hatano}$^1$
         and
    Minoru {\sc Takahashi}
 \vspace{5 mm}\\
        {\it Institute for Solid State Physics, University of Tokyo,}\\
        {\it Roppongi, Minato-ku, Tokyo 106} \\
        {\it ${}^{ 1}$Department of Physics, University of Tokyo,}\\
        {\it Hongo, Bunkyo-ku, Tokyo 113,} \\
        {\it and Department of Physics, Harvard University,}\\
        {\it Cambridge, MA 02138, USA  } \\
\vspace{3mm}
\end{center}

(Received    \hspace{5cm})
%
\begin{abstract}
The finite-size scaling function of the nonlinear susceptibility
of the ferromagnetic Heisenberg chain is given explicitly.
It is conjectured that the scaling function is universal for any values of $S$.
The conjecture is based on the exact solution of the nonlinear susceptibility
for $S=\infty$, and numerical calculations for $S=1/2$ and $S=1$.

\vspace{2mm}
\end{abstract}
\pagebreak
%
\section{Introduction}\label{intro}
\setcounter{equation}{0}
\newcommand{\qbSpin}{\mbox{\boldmath $S$}}
\newcommand{\cbSpin}{\mbox{\boldmath $n$}}
\newcommand{\no}{\nonumber}
\newcommand{\pe}{\mbox e}

Thanks to recent development of technology, it has become possible to
synthesize many quasi-low-dimensional systems experimentally.
This has led to growing theoretical interest in low-temperature properties of
low-dimensional systems.
A remarkable example is the Haldane conjecture\cite{Haldane83a,Haldane83b}
for the spin-$S$ antiferromagnetic Heisenberg chain.

Recently, two of the present authors considered\cite{NT} low-temperature
low-magnetic-field properties of the spin-$S$ ferromagnetic Heisenberg chain.
The Hamiltonian of the model is defined by
\begin{equation}
{\cal H} = - \frac{J}{S^2} \sum_{i=1}^{L}
            {\qbSpin}_{i}\cdot {\qbSpin}_{i+1}
           - \frac{h}{S} \sum_{i=1}^{L}
              S_{i}^{z} ,  \label{eq.1.1}
\end{equation}
where
the coupling constant $J$ is positive and $h$ denotes a magnetic field.
They calculated\cite{NT} the free energy and the magnetization of the
$S=1/2$ system numerically, employing the thermal
Bethe-ansatz method\cite{Takahashi} for the thermodynamic limit $L\to\infty$.
They found\cite{NT} that the magnetization for the $S=1/2$ system has the same
scaling function as
that of the classical ($S=\infty$) ferromagnetic Heisenberg chain:
\begin{eqnarray}
m(T,h) &\simeq& \tilde m_0 \left( x \right) ,  \label{eq.1.100} \\
	 &=& \frac{2}{3} x - \frac{44}{135} x^3
+{\rm O} \left( x^5 \right).  \label{eq.1.110}
\end{eqnarray}
Here $x$ is the scaling variable:
\begin{equation}\label{eq:scaling-x}
x\equiv \frac{Jh}{T^2}.
\end{equation}
They thereby conjectured that this scaling function
should be universal, or common to all values of $S.$
This conjecture implies that the spin-wave excitation from
the ground state of the ferromagnet show universal behavior.

In order to confirm this conjecture for $1\leq S<\infty,$ we have to obtain
information from numerical calculations of finite systems, because
the Bethe ansatz method is not applicable to higher-spin Heisenberg chains.
In the previous paper,\cite{NHT}
we hence generalized the above scaling form
(\ref{eq.1.100}) to finite-size
scaling
and conjectured that the scaling function of the magnetization,
\begin{equation}
 m(T, h, L) \simeq    \tilde m
    \left( x ,  y    \right),  \label{eq.1.conj}
\end{equation}
is universal for
the arbitrary-$S$ ferromagnetic Heisenberg chain.
Here the scaling parameters $x$ and $y$ are defined by (\ref{eq:scaling-x}) and
\begin{equation}\label{eq:scaling-y}
y \equiv \frac{J}{TL}.
\end{equation}

It is difficult to obtain the magnetization $m(T,h,L)$ for the classical
system. Hence,
as the first step, we treated\cite{NHT} the finite-size scaling form of the
linear susceptibility.
The scaling form (\ref{eq.1.conj}) is followed by
the finite-size scaling function of the linear susceptibility $\chi_1$ in the
form
\begin{equation}
\chi_1 (T,L) \equiv  \left. \frac{ \partial m }{ \partial h }
\right|_{h=0} \simeq \frac{J}{T^2} \tilde{\chi}_1 \left( \frac{J}{TL} \right),
\label{eq.1.120}
\end{equation}
where
\begin{equation}
 \tilde \chi_1\left( y \right) \equiv
          \left. \frac{ \partial {\tilde m (x,y) } }{ \partial x }
          \right|_{x=0} . \label{eq.1.130}
\end{equation}
In the previous paper,\cite{NHT}
we analytically obtained the function $\tilde \chi_1$ for the classical,
or the $S=\infty$ Heisenberg chain
with both periodic and open boundary conditions.
Moreover, we showed
that the function $\tilde\chi_1$ for $S=\infty$ fits numerical data for
$S=1/2$ and $S=1$ quite well.
Thus we partly confirmed the conjecture on the universality of the
scaling function (\ref{eq.1.conj}).

In the present paper, we further treat the scaling function of the
nonlinear susceptibility, and show its universality.
Equation (\ref{eq.1.conj}) ensures the universality not only
of the linear susceptibility $\chi_1$ but also of the third-order
nonlinear susceptibility $\chi_3$:
\begin{equation}
\chi_{3} (T,L) \equiv  \left. \frac{ 1}{3 !}
 \frac{ \partial^3 m }{\partial h^3 }   \right|_{h=0} \simeq
\frac{J^3}{T^6} \tilde \chi_3\left(\frac{J}{TL} \right),  \label{eq.1.200}
\end{equation}
where
\begin{equation}
 \tilde \chi_3 \left( y \right) \equiv
          \left.
 \frac{1}{3 !}
 \frac{ \partial^3 \tilde m (x,y) }{ \partial x^3 }
  \right|_{x=0}  . \label{eq.1.210}
\end{equation}
The result (\ref{eq.1.110}) in the thermodynamic limit requires the relation
\begin{equation}
{\tilde \chi_3} (y=0) = - \frac{44}{135} . \label{eq.1.4}
\end{equation}
In the present paper
we obtain an analytical expression of $\tilde \chi_3$ for the classical chain.
(This was partially reported in the previous paper.\cite{NHT})
We then show that the function $\tilde \chi_3$ obtained for $S=\infty$ is
consistent with numerical data for quantum systems,
namely, $S=1/2$ and $S=1$.
Thus we confirm the universality of the scaling function (\ref{eq.1.conj}) up
to the third order of $x$.

In the course of the study, we also obtain the exact expressions of the
four-point correlations and the nonlinear susceptibility of the classical
system.
To our knowledge, these expressions for the periodic classical system
have not appeared in the literature.


Before going into details of calculations,
we comment here on the definition of the scaling limit.
We see in the following that there are corrections of the form $y\sqrt{T}$,
$yT$, {\em etc}.\ to the finite-size scaling functions
(\ref{eq.1.120}) and (\ref{eq.1.200}).
These are less singular than the leading term of $y$
for large systems at low temperatures.
We have to exclude these correction terms in order to obtain the scaling
function of $y$.
We hence extrapolate the limit $T\to0$ while {\em fixing} the scaling variable
$y=J/(LT)$.
We refer to this limit as the {\em scaling limit}.
It should not be confused with the thermodynamic limit $y\to0$, or the limit
$L\to\infty$ with $T$ fixed.
Even after taking the scaling limit, we still consider a finite
value of $y=J/(TL)$.

We organize the paper as follows.
In \S2 we analytically calculate the finite-size scaling function of $\chi_3$
of the classical Heisenberg model for the periodic and open
boundary conditions.
We show numerical confirmation of the universality of the scaling function
for $S=1/2$ and $S=1$ in \S3.
In \S4
we discuss origins of the universality of
the third-order nonlinear susceptibility.
In Appendix A we present in detail the
analytic calculation of the four-point correlation
function
of the classical Heisenberg chain
with periodic boundary condition.
In Appendix B we show the derivation of the third-order nonlinear
susceptibility of the classical Heisenberg chain.

\section{Finite-Size Scaling Functions of
the Third-Order Nonlinear Susceptibility}\label{sec.2}
\setcounter{equation}{0}
In the previous paper\cite{NHT} we presented the finite-size scaling function
of the linear susceptibility,
$\tilde \chi_1 $.
We used the classical Heisenberg chain to obtain the analytic form of
$\tilde \chi_1.$
In this section
we derive the finite-size scaling function
of the third-order nonlinear susceptibility, $\tilde \chi_3 ,$
analytically for the classical Heisenberg chain
with the periodic and the open boundary conditions.
We insist that
the ferromagnetic Heisenberg chain has this finite-size scaling function
$\tilde \chi_3$
not only in the classical case ($S=\infty$)
but also in the quantum case.

The classical Heisenberg chain is given by
\begin{eqnarray}
 {\cal H} &=& - \frac{J}{S^2}  \sum_{i=1}^{L} { \qbSpin }_{i}
\cdot { \qbSpin }_{i+1} -   \frac{h}{ S} \sum_{i=1}^{L} { S}_{i}^{z}
\no \\
        &\rightarrow&  - J \sum_{i=1}^{L}
        \cbSpin_{i} \cdot \cbSpin_{i+1}
        - h \sum_{i=1}^{L} n_{i}^{z}  \quad \mbox{as} \quad
S \rightarrow \infty, \label{eq.2.1}
\end{eqnarray}
where \{ $  {\cbSpin}_{i} $ \} are vectors of length unity.
The nonlinear susceptibility is given by summation of four-point correlations
as follows:
\begin{eqnarray}
\chi_3 (T,L) &=& \frac{1}{3! LT^3 S^4 } \sum_{i_1,i_2,i_3,i_4}
     \left[
     \left\langle
      S_{i_1}^{z}
      S_{i_2}^{z}
      S_{i_3}^{z}
      S_{i_4}^{z}
     \right\rangle
     \right.     \nonumber \\
      & & -
     \left\langle
      S_{i_1}^{z}
       S_{i_2}^{z}
     \right\rangle
     \left\langle
      S_{i_3}^{z}
      S_{i_4}^{z}
     \right\rangle             \nonumber \\
      & & -
     \left\langle
      S_{i_1}^{z}
      S_{i_3}^{z}
     \right\rangle
     \left\langle
       S_{i_2}^{z}
      S_{i_4}^{z}
     \right\rangle             \nonumber \\
      & & -
     \left.
     \left\langle
      S_{i_1}^{z}
      S_{i_4}^{z}
     \right\rangle
     \left\langle
      S_{i_3}^{z}
      S_{i_2}^{z}
     \right\rangle
     \right]  , \label{eq.1.5}
\end{eqnarray}
where $ \langle \cdots \rangle$  denotes the thermal average.

\subsection{Periodic boundary condition}
First,
we consider the periodic system.
To obtain the nonlinear susceptibility, we have to calculate the
partition function, the two-point correlation function and the four-point
correlation function.

Joyce\cite{Joyce} obtained
the partition function of the periodic chain, $Z_L,$
in the form
\begin{equation}
Z_L (K) = \sum_{l=0}^{\infty} (2l+1) \lambda_{l} (K)^L ,
\label{class.prt}
\end{equation}
where $K \equiv J/T$ and
\begin{equation}
\lambda_{l} (K) \equiv  \sqrt{ \frac{ \pi }{2K} }
\ \ {I}_{l+ \frac{1}{2} } (K) \label{eq.2.8}
\end{equation}
with $ {I}_{l+ \frac{1}{2} } (K) $
being modified Bessel functions of the first kind.
The two-point correlation function schematically shown in Fig.\ \ref{f.1}
is given by\cite{Joyce,NHT}
\begin{equation}
\langle  n_{1}^{z} n_{ 1+M }^{z} \rangle_L^{\rm peri}
       = \frac{1}{ 3 Z_L }
         \sum_{l=0}^{\infty} (l+1) \lambda_{l}^{L}
         \left[
            u_l^{M}
            +
            u_l^{L-M}
         \right] , \label{eq.2.17}
\end{equation}
where the bracket $\langle\cdots\rangle_L^{\rm peri}$ denotes the thermal
average for the periodic system of length $L$, and the function $u_l$ is
defined by
\begin{equation}
u_l(K) \equiv   \frac{ \lambda_{l+1}(K) }{ \lambda_{l}(K) }   .\label{eq.2.11}
\end{equation}
 From the calculations given in Appendix A, we also obtain
the four-point correlation function shown in Fig.\ \ref{f.2} as
\begin{eqnarray}
\lefteqn{ \langle  n_{1}^{z} n_{ 1+A }^{z}
	   n_{1+A+B }^{z} n_{ 1+A+B+C }^{z}  \rangle_L^{\rm peri} } \no \\
 &=&  \frac{1}{ Z_L }
    	\sum_{l=0}^{\infty} \lambda_l^L
      \left\{ f_l \left[
         u_l^A v_l^B u_l^C   	+
         u_l^B v_l^C u_l^{D}
        + u_l^C v_l^{D} u_l^A	+
         u_l^{D} v_l^A u_l^B \right]  \right.
        \no \\
  & & \left.   + g_l \left[ u_l^A u_l^C + u_l^B u_l^{D} \right]
	        \right\} ,
	\label{eq.2.18}
\end{eqnarray}
where $D=L-A-B-C$,
\begin{eqnarray}
 f_l &\equiv& \frac{2}{15} \frac{(l+1)(l+2)}{2l+3}
			 ,\label{eq.2.19} \\
 g_l &\equiv& \frac{1}{15} \frac{(l+1)(4l^2+8l+5)}{(2l+1)(2l+3)}
		 ,\label{eq.2.20}
\end{eqnarray}
and
\begin{equation}
v_l(K) \equiv   \frac{ \lambda_{l+2}(K) }{ \lambda_{l}(K) }  .\label{eq.2.12}
\end{equation}

These expressions (\ref{class.prt})-(\ref{eq.2.12}) give the exact expression
of the nonlinear susceptibility (\ref{eq.1.5}).
See Appendix B for details.
Now we calculate its scaling form.

In the scaling limit, that is, in the limit
$T\rightarrow0$ with $y=J/(TL)$ fixed,
the partition function is expressed in terms of $y$ as follows:\cite{NHT}
\begin{eqnarray}
\lefteqn{
\frac{ Z_L (K) }{ \lambda_0(K)^L }
=\sum_{l=0}^\infty (2l+1)\left(\frac{\lambda_l}{\lambda_0}\right)^L
}\nonumber\\
&\rightarrow&
  {\tilde {\cal W}}(y)
\equiv \sum_{l=0}^{\infty}
 (2l+1) \exp \left\{-\frac{l(l+1)}{2y} \right\}.
	\label{eq.2.10}
\end{eqnarray}
Next, we can obtain
the finite-size scaling function (\ref{eq.1.200})
of the third-order susceptibility with the
periodic boundary condition,
${\tilde \chi_3^{\rm peri} },$
as follows:
%
%
\begin{eqnarray}
 {\tilde \chi_3^{\rm peri} } (y) &=&
 \lim_{T\rightarrow 0 \atop {\mbox{\scriptsize $y$ fixed}}}
\left[ \frac{\chi_3^{\rm peri}(T,L)  T^6}{J^3}  \right]  \no \\
  &=& \frac{1}{ \tilde {\cal W} (y) }
  	\sum_{l=0}^{\infty}
  	\exp\left[ - \frac{l(l+1)}{2y} \right]
	\left\{ \frac{8}{15} \frac{l+2}{2l+3}
  	\right.
  	 \no \\
  & & \left.
  	   \times
  	   \left[ -\frac{1}{(l+2)y} \exp \left(-\frac{ l+1}{y} \right)
           \right.
     \right.
     \no \\
  & & \left.
           \left.
            + \frac{1}{(2l+3)(l+1)}
  	       \left( 1- \exp \left(-\frac{l+1}{y}\right) \right)
  	   \right.
  	\right.
 	\no \\
  & & \left.
  	  \left.
  	       - \frac{l+1}{(2l+3)(l+2)^2}
  	       \left( \exp \left(-\frac{2l+3}{y} \right)
  	          - \exp\left(-\frac{l+1}{y} \right)
     	       \right)
     	    \right]
     	 \right. \no \\
  & &  \left.
          + \frac{2}{15}  \frac{4l^2+8l+5}{(2l+1)(2l+3)}
     	 \right. \no \\
  & &  \left.
  	    \times
   	    \left[
         \frac{1}{(l+1)y} \left( 1+\exp \left( -\frac{l+1}{y} \right) \right)
            \right.
       \right. \no \\
 & & \left.
      \left.
      - \frac{2}{(l+1)^2}
   	      \left(
   	         1 - \exp\left(-\frac{l+1}{y} \right)
   	      \right)
   	    \right]
   	 \right\}  \no \\
  & &  -  \frac{1}{ {\tilde {\cal W} (y) }^2} \frac{2}{9y}   . \label{eq.2.21}
\end{eqnarray}

The key in the calculations of the above scaling forms is to note the
behavior of the functions $u_l$ and $v_l$.
These functions behave at low temperatures as follows:\cite{NHT}
\begin{eqnarray}
u_l(K) &\rightarrow& 1 - \frac{l+1}{K} +O\left( \frac{1}{K^2} \right),
\label{eq.2.13} \\
v_l(K) &\rightarrow& 1 - \frac{2l+3}{K} +O\left( \frac{1}{K^2}
\right).\label{eq.2.14}
\end{eqnarray}
Hence they converge to unity as $T\to0$, or $K=J/T\to\infty$
except for the terms $u_l^L$ and $v_l^L$.
Using the scaling parameter $y=K/L,$
we have the following asymptotic forms
of $u_l^L$ and $v_l^L$ in the scaling limit:\cite{NHT}
\begin{eqnarray}
u_l(K)^L
	   &=&\left[ 1-\frac{l+1}{y} \frac{1}{L}
	   			+ O\left( \frac{1}{L^2} \right) \right]^L \no \\
	   &\to&  \exp\left( -\frac{l+1}{y} \right), \label{eq.2.15} \\
v_l(K)^L
	   &=&\left[ 1-\frac{2l+3}{y} \frac{1}{L}
	   			+ O\left( \frac{1}{L^2} \right) \right]^L \no \\
	   &\to&  \exp\left( -\frac{2l+3}{y} \right). \label{eq.2.16}
\end{eqnarray}
Further details are given in Appendix B.

In the thermodynamic limit $y\rightarrow0,$
the right-hand side of eq.\ (\ref{eq.2.21}) is reduced to
\begin{equation}
\tilde \chi_3^{\rm peri} (y\rightarrow 0) \rightarrow -\frac{44}{135}
\label{eq.2.22}
\end{equation}
This value $-44/135$ is consistent with the value in the thermodynamic limit,
eq.\ (\ref{eq.1.4}).

\subsection{Open boundary condition}
Next we consider the open system of length $L.$
Because we handle only three modes $l=0,1,2$ of the modified Bessel functions
in the open systems,
it is easier to calculate the correlation functions in the open systems
than in the periodic systems.
The partition function is identical to unity.
Fisher\cite{Fisher} gave the two-point function as follows:
\begin{equation}
\langle n_{1}^z n_{1+M}^z \rangle^{\rm open}
= \frac{1}{3} u_0 (K)^M   . \label{eq.2.23}
\end{equation}
Tomita and Mashiyama\cite{TM} gave the four-point function as  follows:
\begin{equation}
\langle n_{1}^z n_{1+A}^z n_{1+A+B}^z n_{1+A+B+C}^z \rangle^{\rm open}
= \frac{1}{9} u_0(K)^A \left[ \frac{4}{5} v_0(K)^B +1 \right]
   u_0 (K)^C .   \label{eq.2.24}
\end{equation}
Using these expressions, we already reported\cite{NHT}
the finite-size scaling function of the third-order susceptibility
with the open boundary condition for the classical Heisenberg chain as follows:
\begin{eqnarray}
{\tilde \chi_3}^{\rm open}(y) &=&
      -       \frac{44}{135}
      -       \frac{32}{45} \exp\left( -\frac{1}{y} \right)
      +      \frac{2y}{405}
\left[1-\exp\left( -\frac{1}{y} \right) \right] \no \\
 & & \ \times \left[169+43\exp\left(-\frac{1}{y}\right)
-2\exp\left(-\frac{2}{y}\right)\right] .
 \label{tchi3}
\end{eqnarray}
In the thermodynamic limit
$y\rightarrow 0,$
we have the value of the finite-size scaling function of the third-order
susceptibility as
\begin{equation}
\tilde \chi_3^{\rm open} (y\rightarrow0) \rightarrow - \frac{44}{135}
.\label{open.chi30}
\end{equation}
This value is consistent with the eq.\ (\ref{eq.1.4}) as well as in
the periodic case (\ref{eq.2.22}).

\section{Finite-Size Calculation for $S=1/2$ and $S=1$}
\setcounter{equation}{0}
In the previous section,
we obtained the finite-size scaling function
$\tilde \chi_3$ for the classical systems
with the periodic and open boundary conditions.
We conjecture\cite{NT,NHT} that
this scaling function is common to
the ferromagnetic Heisenberg chain with an arbitrary magnitude of the spin.
In this section,
in order to check this conjecture,
we calculate the third-order susceptibility (\ref{eq.1.5}) of finite systems
with
$S=1/2$ and $S=1$ by the Householder method
and compare the numerical data in the scaling limit
with the finite-size scaling function (\ref{eq.2.21}) and (\ref{tchi3}).

\subsection{Periodic boundary condition}
First,
we treat the periodic system.
We diagonalized the Heisenberg Hamiltonian (\ref{eq.1.1})
numerically by the Householder method.
The system size is up to $L=14$ for $S=1/2$
and up to $L=10$ for $S=1$.
We then calculated the spin correlation functions
and summed them up over all sites following eq.\ (\ref{eq.1.5}).
We thus obtained the third-order nonlinear susceptibility
numerically.

We plot the numerical data
of $S=1/2$ chains with the periodic boundary condition in Fig.\ \ref{f.5}
and those of $S=1$ chain with the periodic boundary condition in
Fig.\ \ref{f.6}.
We observe corrections to finite-size scaling in these figures.
In order to take the scaling limit, or to exclude the corrections,
we here assume that
the leading correction is of the same form as in the linear
susceptibility:\cite{NHT}
\begin{equation}
\chi_3 (T,L) = \frac{J^3}{T^6} \left[ \tilde \chi_3 (y) + O(\sqrt{T/J})
\right].  \label{corr.chi3}
\end{equation}

In the previous paper\cite{NHT} we take the scaling limit, fitting the data
quadratically with respect to $\sqrt{T/J}$.
In the present analysis, however, the quadratic fitting resulted in spurious
scaling limit.
Some of the extrapolation curves for the quadratic fitting have the maximum
around $\sqrt{T/J}\sim0.2$ and start decreasing as $T/J\to0$.
This indicates that corrections higher than the second order are large
in the nonlinear susceptibility.
We therefore
used corrections to finite-size scaling up to the third-order
term, i.e.\ $(T/J)^{3/2}$.
Each fitting curve in Figs.\ \ref{f.5} and \ref{f.6} was
determined by the four points nearest to the ordinate.
The crosses on the ordinate of Figs.\ \ref{f.5} and \ref{f.6}
denote the scaling limit $\tilde \chi_3^{\rm peri} (y)$.

In Fig.\ \ref{f.7},
we summarize the data in the scaling limit obtained in Figs.\ \ref{f.5} and
\ref{f.6} together with
the finite-size scaling function for the classical case (\ref{eq.2.21}).
(In calculating eq.\ (\ref{eq.2.21}) we numerically summed the terms up to
$l=100$, which turned out to be sufficient for the convergence of the series.)
The data for $S=1/2$ and $S=1$ are quite consistent with
the finite-size scaling function for $S=\infty.$
This suggests that
the finite-size scaling function (\ref{eq.2.21}),
which is given analytically for the classical chain,
does not depend on the magnitude of the spin $S.$

\subsection{Open boundary condition}
Next,
we treat the open chain.
We perform in Figs.\ \ref{f.8} and \ref{f.9} the same analysis as in the
periodic systems, i.e.\ Figs.\ \ref{f.5} and \ref{f.6}.
The finite-size scaling function (\ref{tchi3}) and the scaling-limit
data from Figs.\ \ref{f.8} and \ref{f.9} are summarized in Fig.\ \ref{f.10}.
We conclude that the finite-size scaling function (\ref{tchi3})
is common to all magnitudes of the spin for the open systems as well as
for the periodic systems.

\section{Discussion}
\setcounter{equation}{0}
By using the numerical calculations for the quantum case, namely $S=1/2$ and
$S=1$,
we showed that the finite-size scaling function $\tilde \chi_3$
derived from the classical case
is universal with respect to the magnitude of the spin
both for the periodic and the open
boundary conditions.
The scaling function $\tilde \chi_3$ is exactly given by
eq.\ (\ref{eq.2.21}) and eq.\ (\ref{tchi3})
for the periodic and the open chain, respectively.
We emphasize that the universality both of the linear and the
third-order susceptibilities implies
the universality (\ref{eq.1.conj}) of the magnetization of the
arbitrary-spin ferromagnetic Heisenberg chain.

The universality of the scaling function implies that the spin-wave excitation
in the present systems show universal behavior.
The ground state of the arbitrary-$S$ Heisenberg ferromagnet in a magnetic
field is obviously the state where all the spins are aligned in the direction
of the field.
The thermal fluctuation creates spin-wave excitations.
The present universality indicates that the thermally excited
spin wave behaves
independently of $S$.
We can see this universal behavior as
the universality of the correlation functions.
We have already suggested\cite{NHT} that the universality
of the linear susceptibility originates
in the universality of the two-point correlation function.
We naturally expect that
the universality of the nonlinear susceptibility originates
in the universality of the two-point and the four-point correlation functions.

Indeed, we can derive the scaling forms
of the correlation functions as follows.
The finite-size scaling functions for the linear susceptibility and the
nonlinear susceptibility are rederived from these scaling forms below.
When the distance between two spins is large,
we have the scaling function of
the two-point correlation function for the periodic chain as follows:\cite{NHT}
\begin{eqnarray}
\lefteqn{ \left\langle
\left( \frac{ S_{1}^{z}}{S} \right)
\left( \frac{ S_{1+M}^{z}}{S} \right)
\right\rangle_L^{\rm peri} }\no \\
&\sim&   \frac{1}{\tilde {\cal W}}
   \displaystyle
          \sum_{l=0}^{\infty}
            \left\{ (l+1) \exp{ \left[ - \frac{l(l+1)}{2y}
  \right] } \right\}
\no\\
&&\times
                        \left[
                          \exp{ \left( -\frac{M}{ \xi_{l}} \right) }
+  \exp{\left( -\frac{L-M}{ \xi_{l}} \right) }
                        \right]
\label{pcor}
\end{eqnarray}
for ${}^{ \forall }S$ and $M,(L-M) \gg 1$,
where the correlation length $\xi_l$ of the mode $l$ is defined by
\begin{equation}
\xi_l \equiv \frac{K}{l+1} . \label{xi}
\end{equation}
We also obtain from eq.\ (\ref{eq.2.18})
the scaling from of the four-point
correlation function for the periodic chain as follows:
\begin{eqnarray}
\lefteqn{  \left\langle
\left( \frac{ S_{1}^{z}}{S} \right)
\left( \frac{ S_{1+A}^{z}}{S} \right)
\left( \frac{ S_{1+A+B}^{z}}{S} \right)
\left( \frac{ S_{1+A+B+C}^{z}}{S} \right)
\right\rangle_L^{\rm peri} } \no \\
&\sim&
     \frac{1}{\tilde {\cal W}}
        \displaystyle
          \sum_{l=0}^{\infty}
          \exp \left[-\frac{ l(l+1)  }{ 2  y }   \right]
\no \\
& &	\times
    \left\{
     f_l
    \left[
	\exp\left( - \frac{ A }{ \xi_l } \right)
	\exp\left( -  \frac{ B }{  \xi_l} \right)
	\exp\left( -  \frac{  B }{\xi_{l+1} } \right)
	\exp\left( -  \frac{ C }{ \xi_l } \right)
  \right.
  \right.
 \no \\
& & \left.
+
	\exp\left( -  \frac{ B }{ \xi_l } \right)
	\exp\left( -  \frac{ C }{  \xi_l} \right)
	\exp\left( -  \frac{ C  }{\xi_{l+1} } \right)
	\exp\left(-  \frac{ D }{  \xi_l } \right)
  \right.
  \no \\
& &	\left. +
	\exp\left( -  \frac{ C }{  \xi_l } \right)
	\exp\left( -  \frac{ D }{  \xi_l} \right)
	\exp\left( -  \frac{ D  }{\xi_{l+1} } \right)
	\exp\left(-  \frac{ A }{ \xi_l } \right)
   \right. \no \\
& & \left.+
	\exp\left( - \frac{ D }{ \xi_l } \right)
	\exp\left( -  \frac{ A }{  \xi_l} \right)
	\exp\left( -  \frac{  A }{\xi_{l+1} } \right)
	\exp\left(-  \frac{ B }{ \xi_l } \right)
  \right]
 	\no \\
 & & +
     \left.
      g_l
	  \left[
		\exp\left( -\frac{B}{\xi_l} \right)
		\exp\left( -\frac{D}{\xi_l} \right)
   +  	\exp\left( -\frac{A}{\xi_l} \right)
		\exp\left( -\frac{C}{\xi_l} \right)
	   \right]
	\right\}
 \label{eq.4.100}
\end{eqnarray}
for ${}^\forall S$ and $A,B,C,D \gg 1$ with $A+B+C+D=L$.

For the open chain,
we have obtained the scaling function of the two-point
correlation function as\cite{NHT}
\begin{equation}
\left\langle
\left( \frac{ S_{1}^{z}}{S} \right)
\left( \frac{ S_{1+M}^{z}}{S} \right)
\right\rangle^{\rm open}  =
\frac{1}{3} \exp{ \left( -M/\xi_0 \right) }   \label{eq.4.110}
\end{equation}
for ${}^{ \forall } S$ and $M \gg 1$.
We may also obtain the scaling function of the four-point
correlation function as
\begin{eqnarray}
\lefteqn{  \left\langle
\left( \frac{ S_{1}^{z}}{S} \right)
\left( \frac{ S_{1+A}^{z}}{S} \right)
\left( \frac{ S_{1+A+B}^{z}}{S} \right)
\left( \frac{ S_{1+A+B+C}^{z}}{S} \right)
\right\rangle^{\rm open} } \no \\
&\sim&
 \frac{1}{9} \exp{ \left( -\frac{A}{\xi_0} \right) }
         \left[ \exp{\left(  -\frac{B}{\xi_0}  \right)} \exp{ \left(
-\frac{B}{\xi_1} \right) } +1 \right]
          \exp{ \left( -\frac{C}{\xi_0}  \right)}  \label{eq.4.120}
\end{eqnarray}
for ${}^{ \forall } S$ and $A,B,C \gg 1$.

\section*{Acknowledgments}
The authors are grateful to
Dr.\ Nishimori for providing the numerical package TITPACK version 2, and to
Dr.\ Kaburagi and Dr.\ Tonegawa for the package KOBEPACK version 1.0.
This work was supported in part by
Grant-in-Aid for Scientific Research on
Priority Areas,
"Molecular Magnetism" (Area No 228) and "Infinite analysis" (Area No 231)
>from the Ministry of Education, Science and Culture.
Calculations which needed a large amount of the computer memory
were done in the FACOM VPP500 of the Supercomputer Center,
the Institute for Solid State Physics,
the University of Tokyo.
\eqreset
\setcounter{section}{0}
\renewcommand{\theequation}{\Alph{section}.\arabic{equation}}
\renewcommand{\thesection}{Appendix \Alph{section}:}

\section{Four-point Correlation Function of the Classical Heisenberg Chain
with the Periodic Boundary Condition}\label{ap.1}
In this Appendix
we show the derivation of the four-point correlation function (\ref{eq.2.18}).

We consider the classical Heisenberg chain (\ref{eq.2.1}) with the
periodic boundary condition.
We choose the direction of the magnetic field $ {\bf h} (| {\bf h} | =h)$
as the $z$ axis of the three-dimensional spin space.
We define the polar coordinate
$\theta_i$ and $\Phi_i$
of the $i$th classical spin relative to the
$z$ axis.
We express each spin in the form
\begin{eqnarray}
n_i^{z} &=& \cos \theta_i  \no \\
	  &=& \sqrt{ \frac{4\pi}{3} } {Y}_{10} ( \theta_i,\Phi_i ) \no \\
	  &\equiv& \sqrt{  \frac{4\pi}{3} } {Y}_{10} (i), \label{a.1}
\end{eqnarray}
where $Y_{lm}$ is the spherical harmonics.
The exponential function of $\cbSpin_i \cdot\cbSpin_{i+1}$ is
expanded in terms of the spherical harmonics as follows:\cite{Joyce}
\begin{eqnarray}
\pe^{K \cbSpin_i \cbSpin_{i+1} } &=& \pe^{ K \cos \Theta} \no \\
	&=& 4\pi \sum_{l=0}^{\infty} \sum_{m=-l}^{l}
		\lambda_l  Y_{l,m}(i) Y_{l,m}^{\ast} (i+1) ,\label{ker}
\end{eqnarray}
where the parameter $\Theta$ is the angle between the spins $\cbSpin_i$and $
\cbSpin_{i+1}$, and $\lambda_l$ is defined by eq.\ (\ref{eq.2.8}).

In the following algebra, we use two formulas concerning the integration
of the spherical harmonics.
First, the integral of the product of
two spherical harmonics over the solid angle $\Omega_i$ is given by\cite{Joyce}
\begin{equation}
\int_0^{2\pi} \int_0^{\pi} {Y}_{l,m}^{\ast}(i)
{Y}_{l^{\prime},m^{\prime} } (i) {\rm d} \Omega_i
= \delta_{l,l^{\prime} } \delta_{m,m^{\prime}} , \label{a.3.1}
\end{equation}
where $\delta$ is the Kronecker symbol.
Next, the integral of the product of three spherical harmonics can be
expressed in terms of Wigner's $3n-j$ symbols:\cite{Joyce,Landau}
\begin{eqnarray}
\lefteqn{ \int_0^{2\pi} {\rm d }\Phi \int_{0}^{\pi}  \sin \theta {\rm d} \theta
Y_{l_1,m_1}^{*} (\theta,\Phi)
Y_{l_2,m_2}(\theta,\Phi)
Y_{l_3,m_3}(\theta,\Phi)
} \no \\
&=& (-1)^{m_1}
\left[ \frac{(2l_1+1)(2l_2+1)(2l_3+1)}{4\pi} \right]^{1/2}
\no \\
&&\times
	\left(
		\begin{array}{c c c}
		l_1 & l_2 & l_3 \\
		0   & 0 & 0
		\end{array}
	\right)
	\left(
		\begin{array}{c c c}
		l_1 & l_2 & l_3 \\
		-m_1  & m_2 & m_3
		\end{array}
	\right) , \no \\
& & 	\label{3j.2}
\end{eqnarray}
where Wigner's $3n-j$ symbols can be written as follows:
\begin{eqnarray}
\lefteqn{	\left(
		\begin{array}{c c c}
		l_1 & l_2 & l_3 \\
		m_1   & m_2 & m_3
		\end{array}
	\right)} \no \\
 &\equiv&
	\delta_{m_1+m_2+m_3,0}
	\left[ \frac{ (l_1+l_2-l_3)! (l_1-l_2+l_3)! (-l_1+l_2+l_3)! }{
	      (l_1+l_2+l_3+1)! }
	\right]^{\frac{1}{2} }  \no \\
 & &	\times
	\left[
	(l_1 +m_1 )! (l_1-m_1)!
	(l_2 + m_2 )!(l_2-m_2)!
	(l_3 + m_3)!(l_3 - m_3)!
	\right]^{\frac{1}{2}} \no \\
 & &	\times
 	\sum_{z} \frac{ (-1)^{z+l_1-l_2-m_3}  }{
 	       z! (l_1+l_2-l_3-z)! (l_1-m_1-z)! (l_2+m_2-z)!
 	       (l_3-l_2+m_1+z)!(l_3-l_1-m_2+z)! } . \no \\
  & &	         \label{3j.def}
\end{eqnarray}
The summation in eq.(\ref{3j.def}) is over all integers $z.$

Using eqs.\ (\ref{a.1}) - (\ref{3j.2}),
we can write down the four-point correlation function
shown in Fig.\ \ref{f.2} as
\begin{eqnarray}
\lefteqn{ F(A,B,C,D:K) \equiv
Z_L {\left\langle
n_1^z n_{1+A}^z n_{1+A+B}^z n_{1+A+B+C}^z \right\rangle}_L
, }
\label{eq.a.100} \\
 &=&  {\left( \frac{3}{4\pi} \right) }^2 \prod_{i=1}^L
 	\int \frac{ {\rm d} \Omega_i }{ 4 \pi}
       \exp(K {\cbSpin}_i {\cbSpin}_{i+1} )
 	{Y}_{10}(1){Y}_{10} (1+A)\no \\
  & & \times {Y}_{10}(1+A+B){Y}_{10}(1+A+B+C)
 	\no \\
 &=&   {\left( \frac{3}{4\pi} \right) }^2
  	\int {\rm d} \Omega_1 \int {\rm d} \Omega_{1+A}
  	\int {\rm d} \Omega_{1+A+B} \int {\rm d} \Omega_{1+A+B+C}
	\no  \\
 & &
  	\sum_{ \{ l_1 m_1 \} }  \sum_{ \{ l_2 m_2 \} }
  	\sum_{ \{ l_3 m_3 \} } \sum_{ \{ l_4 m_4 \} }
 	\lambda_{l_1}^A \lambda_{l_2}^B \lambda_{l_3}^C \lambda_{l_4}^{D}
 	\no  \\
 & &	\times
 	{Y}_{l_4 m_4}^{\ast}(1)  {Y}_{10}(1) {Y}_{l_1 m_1}(1)
 	\times
 	{Y}_{l_1 m_1}^{*}(1+A)  {Y}_{10}(1+A) {Y}_{l_2 m_2}(1+A)
	\no  \\
 & &	\times
 	{Y}_{l_2 m_2}^{*} (1+A+B) {Y}_{10}(1+A+B) {Y}_{l_3 m_3}(1+A+B)
 	\no  \\
 & &	\times
 	{Y}_{l_3 m_3}^{*} (1+A+B+C) {Y}_{10}(1+A+B+C) {Y}_{l_4 m_4}(1+A+B+C)
 	\no \\
 &=& { \left( \frac{3}{4\pi} \right) }^2
  	\sum_{ \{ l_1 m_1 \} }  \sum_{ \{ l_2 m_2 \} }
  	\sum_{ \{ l_3 m_3 \} } \sum_{ \{ l_4 m_4 \} }
 	\lambda_{l_1}^A \lambda_{l_2}^B \lambda_{l_3}^C \lambda_{l_4}^{D}
   \no  \\
 & &	\times
	(-1)^{m_1+m_2+m_3+m_4}
	\frac{9 (2l_1+1) (2l_2+1) (2l_3+1) (2l_4+1) }{ {(4\pi)}^2 }
 	\no  \\
 & &	\times
	\left(
		\begin{array}{c c c}
		l_4 & 1 & l_1 \\
		0   & 0 & 0
		\end{array}
	\right)
	\left(
		\begin{array}{c c c}
		l_4 & 1 & l_1 \\
		-m_4   & 0 & m_1
		\end{array}
	\right)
	\left(
		\begin{array}{c c c}
		l_1 & 1 & l_2 \\
		0   & 0 & 0
		\end{array}
	\right)
	\left(
		\begin{array}{c c c}
		l_1 & 1 & l_2 \\
		-m_1  & 0 & m_2
		\end{array}
	\right)
 	\no  \\
 & &	\times
	\left(
		\begin{array}{c c c}
		l_2 & 1 & l_3 \\
		0   & 0 & 0
		\end{array}
	\right)
	\left(
		\begin{array}{c c c}
		l_2 & 1 & l_3 \\
		-m_2  & 0 & m_3
		\end{array}
	\right)
	\left(
		\begin{array}{c c c}
		l_3 & 1 & l_4 \\
		0   & 0 & 0
		\end{array}
	\right)
	\left(
		\begin{array}{c c c}
		l_3 & 1 & l_4 \\
		-m_3  & 0 & m_4
		\end{array}
	\right).\no \\
& &
	\label{a.2}
\end{eqnarray}
In order to calculate eq.\ (\ref{a.2}) further,
we need the following formula for the Wigner symbol:\cite{Landau}
\begin{equation}
\left(
	\begin{array}{c c c}
		l & 1 & l^{\prime} \\
		-m  & 0 & m^{\prime}
	\end{array}
\right)
= \delta_{m,m^{\prime}}
\left(
	\begin{array}{c c c}
		l & 1 & l^{\prime} \\
		-m  & 0 & m
	\end{array}
\right) ,\label{formla.1}
\end{equation}
with
\begin{eqnarray}
\left(
	\begin{array}{c c c}
		l & 1 & l^{\prime} \\
		-m  & 0 & m
	\end{array}
\right)
&=&
 	\delta_{l^{\prime},l+1}
 	\left(
	\begin{array}{c c c}
		l & 1 & l+1 \\
		-m  & 0 & m
	\end{array}
	\right)
\no \\
&& +
 	\delta_{l^{\prime},l-1}
 	\left(
	\begin{array}{c c c}
		l & 1 & l-1 \\
		-m  & 0 & m
	\end{array}
	\right)
 +
  	\delta_{l^{\prime},l}
 	\left(
	\begin{array}{c c c}
		l & 1 & l \\
		-m  & 0 & m
	\end{array}
	\right) . \no \\
 & & 		\label{formula.2}
\end{eqnarray}
Each term in eq.\ (\ref{formula.2}) is given as follows:\cite{Landau}
\begin{eqnarray}
\left(
	\begin{array}{c c c}
		l & 1 & l \\
		-m  & 0 & m
	\end{array}
\right)
 &=& (-1)^{-l+m} \frac{m}{\sqrt{ l (l+1) (2l+3) } } , \label{fa.1} \\
\left(
	\begin{array}{c c c}
		l & 1 & l+1 \\
		-m  & 0 & m
	\end{array}
\right)
 &=& (-1)^{l-m+1} \sqrt{ \frac{(l-m+1)(l+m+1)}{(l+1)(2l+1)(2l+3)} }, \no \\
 & & \label{fa.2} \\
\left(
	\begin{array}{c c c}
		l & 1 & l-1 \\
		-m  & 0 & m
	\end{array}
\right)
 &=& (-1)^{l-m} \sqrt{ \frac{(l-m)(l+m)}{l(2l-1)(2l+1)} }. \label{fa.3}
\end{eqnarray}
We substitute the $3n-j$ symbols
in eq.\ (\ref{a.2}) with  eqs.\ (\ref{formla.1})-(\ref{fa.3}).
Then we have the form
\begin{eqnarray}
\lefteqn{ F(A,B,C,D:K) } \no \\
 &=&  \sum_{ \{ l_1 m_1 \} }  \sum_{ \{ l_2 m_2 \} }
  	\sum_{ \{ l_3 m_3 \} } \sum_{ \{ l_4 m_4 \} }
 	\lambda_{l_1}^A \lambda_{l_2}^B \lambda_{l_3}^C \lambda_{l_4}^{D}
\no\\
&&\times
 	(2l_1+1) (2l_2+1) (2l_3+1) (2l_4+1)
 	\no  \\
 & &	\times
	\{ a(l_4 +1, m) \delta_{l_1,l_4+1} + a(l_4 , m) \delta_{l_1,l_4 -1} \}
 	\no  \\
 & &	\times
	\{ a(l_1 +1, m) \delta_{l_2,l_1+1} + a(l_1 , m) \delta_{l_2,l_1 -1} \}
 	\no  \\
 & &	\times
 	\{ a(l_2 +1, m) \delta_{l_3,l_2+1} + a(l_2 , m) \delta_{l_3,l_2 -1} \}
 	\no  \\
 & &	\times
	\{ a(l_3 +1, m) \delta_{l_4,l_3+1} + a(l_3  , m) ,\delta_{l_4,l_3 -1} \},
 	\label{a.3}
 \end{eqnarray}
where
\begin{equation}
a(l,m) \equiv (-1)^m \frac{ \sqrt{ l^2 -m^2 } }{(2l-1) (2l+1) }.
\label{a.4}
\end{equation}
\newcommand{\lgp}{\mbox {\boldmath $l$}}

Because of Kronecker's deltas, the nonvanishing contributions of
the fourfold summation 
in eq.\ (\ref{a.3}) come from the following six combinations:
\begin{eqnarray}
\lgp_{\alpha 1}  	&\equiv & \{ l_1= l, l_2=l+1 , l_3=l+2 ,l_4=l+1 \}
,\label{al.1} \\
\lgp_{\alpha 2 }  	&\equiv & \{ l+1, l, l +1,l+2 \} ,\label{al.2} \\
\lgp_{\alpha 3} 	&\equiv & \{  l+2,l+1 , l,l+1 \}, \label{al.3} \\
\lgp_{\alpha 4}  	&\equiv & \{ l+1, l+2   , l+1 ,l \} ,\label{al.4} \\
\lgp_{\beta 1 }  	&\equiv & \{ l, l+1   , l ,l+1 \} ,\label{al.5} \\
\lgp_{\beta 2 }  	&\equiv & \{ l+1, l   , l+1 ,l \} .\label{al.6}
\end{eqnarray}
Hence, it is sufficient to sum up the terms over the one parameter $l$.
The summation of the contributions from the first four combinations,
$\lgp_{\alpha 1 },
\lgp_{\alpha 2 },\lgp_{\alpha 3 } ,\lgp_{\alpha 4 }$,
and that from the last two combinations, $\lgp_{\beta 1}$, $\lgp_{\beta2}$,
are given as follows, respectively:
\begin{eqnarray}
F^{\alpha}(A,B,C,D: K)
 &\equiv& 	   \sum_{l,m}  (2l+1){(2l+3)}^2(2l+5)
\no \\
& & \times	{a(l+1,m)}^2 {a(l+2,m)}^2 \no \\
& &	\times
  \left(
      \lambda_l^A \lambda_{l+1}^B \lambda_{l+2}^C \lambda_{l+1}^D
   + \lambda_l^B \lambda_{l+1}^C \lambda_{l+2}^D \lambda_{l+1}^A
 \right.
\no \\
&& \left.
   + \lambda_l^C \lambda_{l+1}^D \lambda_{l+2}^A \lambda_{l+1}^B
   + \lambda_l^D \lambda_{l+1}^A \lambda_{l+2}^B \lambda_{l+1}^C
  \right)   \no \\
 &=&	 \sum_{l=0}^{\infty}
	  f_l
	 \lambda_{l}^L
  \left[
	 {u_l  }^B  {v_l}^C  {u_l }^D
+	 {u_l  }^C  {v_l}^D  {u_l }^A
	\right.
\no \\
& &	\left .
+	 {u_l  }^D  {v_l}^A  {u_l }^B
+	 {u_l  }^A  {v_l}^B  {u_l }^C
  \right]
	\label{alpha},
\end{eqnarray}
and
\begin{eqnarray}
F^{\beta}(A,B,C,D :K)
 &\equiv& \sum_{l,m}   {(2l+1)}^2 {(2l+3)}^2 {a(l+1,m)}^4
\no \\
&&\times
\left[
	\lambda_l^A \lambda_{l+1}^B \lambda_l^C \lambda_{l+1}^D +
   \lambda_{l+1}^A \lambda_{l}^B \lambda_{l+1}^C \lambda_{l}^D
\right] \no \\
 &=& \sum_{l=0}^{\infty}
 	         g_l
         \lambda_{l}^L
     \left[
         { u_l}^B
         { u_l}^D
 +       { u_l}^A
         { u_l}^C
      \right]  . \label{beta}
\end{eqnarray}
Here the functions $u_l$ and $v_l,$ and the coefficient $f_l$ and $g_l$
are defined
by eqs.\ (\ref{eq.2.11}), (\ref{eq.2.12}), (\ref{eq.2.19})
and (\ref{eq.2.20}), respectively.

Summing up (\ref{alpha}) and (\ref{beta}),
we arrive at the analytic expression of
the four-point correlation function in the form eq.\ (\ref{eq.2.18}).

\section{The Third-order Nonlinear Susceptibility of the Classical Heisenberg
Chain with the Periodic Boundary Condition}\label{ap.2}
\setcounter{equation}{0}
We write the finite-size scaling function of the third-order nonlinear
susceptibility for the periodic system as eq.\ (\ref{eq.2.21}).
We present a deviation of this finite-size scaling function in this Appendix.

We substitute the correlation functions in the right-hand side of
eq.\ (\ref{eq.1.5}) with  eqs.\ (\ref{eq.2.17}) and (\ref{eq.2.18}).
We divide this summation into the summation of the four-point correlation
functions $\Phi$
and that of the two-point correlation functions $\Psi$ as follows:
\begin{equation}
\chi_3^{\rm peri}(T,L) = \frac{1}{3!LT^3}\left[ \Phi(T,L) - \Psi(T,L) \right] ,
\label{eq.b.100}
\end{equation}
where
\begin{eqnarray}
\Phi(T,L) &\equiv& \sum_{i_1, i_2,i_3,i_4}
               \langle n_{i_1}^z n_{i_2}^z n_{i_3}^z n_{i_4}^z \rangle_L^{\rm
peri} , \label{eq.b.200} \\
\Psi(T,L) &\equiv& 3 \left[ \sum_{i_1, i_2 }
               \langle n_{i_1}^z n_{i_2}^z  \rangle_L^{\rm peri} \right]^2
.\label{eq.b.300}
\end{eqnarray}

Let us show in the following the calculations of $\Phi$ in details.
The fourfold summation over the sites $i_1,i_2,i_3$ and $i_4$
in eq.\ (\ref{eq.b.200}) is
reduced to the threefold summation because of the translational invariance of
the periodic system.
By using the parameters $A,B,C$ and $D$ shown in Fig.\ \ref{f.2},
we express the summation of the four-point correlation function, $\Phi$,
as follows:
\begin{eqnarray}
\Phi(T,L)
  & =& \frac{L}{Z_L} \left[
          \phi_1(T,L) +   \phi_2(T,L) +  \phi_3 (T,L) +  \phi_4(T,L)
       \right. \no \\
   & & \left. +     F( 0,0,0 ,L:K) \right],   \label{eq.b.400}
\end{eqnarray}
where $Z_L$ is the partition function (\ref{class.prt}) ,
$F$ is defined by eq.\ (\ref{eq.a.100}), and
\begin{eqnarray}
 \phi_1(T,L) &\equiv&
       (4-1)! \sum_{1 \le A,B,C,D \le L } \delta_{L,A+B+C+D}  F( A,B,C,D :K),
\no \\
        & &  \label{eq.b.410} \\
 \phi_2(T,L) &\equiv&
      12 \sum_{1 \le B,C,D \le L} \delta_{L,B+C+D}
          F( 0,B,C,D :K)  ,  \label{eq.b.420} \\
 \phi_3(T,L) &\equiv& 4 \sum_{1 \le C,D \le L } \delta_{L,C+D}
        F( 0,0,C,D :K) ,\label{eq.b.430} \\
\phi_4(T,L) &\equiv&  {}_3 {\rm C}_1\sum_{1 \le C,D \le L } \delta_{L,C+D}
       F( 0,C,0,D :K)   . \label{eq.b.440}
\end{eqnarray}

Now we calculate each term of (\ref{eq.b.400}) as follows.
First, using eq.\ (\ref{eq.2.18}),
we write $\phi_1$ explicitly in the form
\begin{eqnarray}
\frac{1}{6} \phi_1 (T,L) &=&
	\sum_{A,B,C,D} \delta_{L,A+B+C+D} \no \\
 & & 	   \times  	\sum_{l=0}^{\infty} \lambda_l^L
      \left\{ f_l \left[
         u_l^A v_l^B u_l^C   	+
         u_l^B v_l^C u_l^{D}
        + u_l^C v_l^{D} u_l^A	+
         u_l^{D} v_l^A u_l^B \right]  \right.
        \no \\
  & & \left.   + g_l \left[ u_l^A u_l^C + u_l^B u_l^{D} \right]
	        \right\}  \no \\
  &=& \sum_{l} \lambda_l^L \left\{
            4 f_l V_1 (K,L)
            + 2 g_l V_2(K,L)
            \right\} ,\label{eq.b.500}
\end{eqnarray}
where we define $V_1$ and $V_2$ as
\begin{eqnarray}
V_1(K,L) &\equiv&
    \sum_{A=1}^{L-3}\ \
    \sum_{B=1}^{L-A-2} \ \
    \sum_{C=1}^{L-A-B-1}
      u_l(K)^A v_l(K)^B u_l(K)^C  , \no \\
   & & 					\label{eq.b.510} \\
V_2(K,L) &\equiv&
    \sum_{A=1}^{L-3} \ \
    \sum_{B=1}^{L-A-2} \ \
    \sum_{C=1}^{L-A-B-1}
      u_l(K)^A  u_l(K)^C  . \label{eq.b.520}
\end{eqnarray}
In obtaining the final expression of (\ref{eq.b.500}), we use the invariance
of the summation of $u_l^A v_l^B u_l^C $ and
$u_l^A u_l^C$ under the cyclic permutation among $A,B,C$ and $D$.
Each summation in (\ref{eq.b.510}) and (\ref{eq.b.520})
can be carried out in the form
\begin{equation}
\sum_{n=1}^Nx^n=\left(1-x^N\right)E(x)
\end{equation}
with
\begin{equation}
E(x) \equiv
\frac{x}{1-x}, \label{eq.b.530}  \end{equation}
for $|x| < 1 .$
We thereby obtain $V_1$ as follows:
\begin{eqnarray}
V_1 &=&  E(u_l)
    \sum_{A=1}^{L-3} \ \
    \sum_{B=1}^{L-A-2}
   \left[ u_l^A v_l^B - u_l^{L-1}
          u_{l+1}^B \right] \no \\
   &=&  E( u_l )
    \sum_{A=1}^{L-3}
   \left\{ E(v_l) \left[ u_l^A -
       v_l^{L-2}
         \left(  \frac{1}{ u_{l+1} } \right)^{A} \right]
  - u_l^{L-1} E\left(
          u_{l+1}\right)
  \left[ 1 -
          u_{l+1}^{L-2-A} \right] \right\} \no \\
 &=& E( u_l )
  \left\{
   -E\left(
           u_{l+1}\right) u_l^{L-1} (L-3)
   +E(u_l)E(v_l) \left( 1 - u_l^{L-3} \right) \right. \no \\
 & & \left.  - \left[ E(v_l) - u_l E\left(
        u_{l+1}\right) \right]
     E\left(
        \frac{1}{ u_{l+1}} \right) v_l \left( v_l^{L-3} - u_l^{L-3} \right)
\right\} .\label{eq.b.540}
\end{eqnarray}
Here we have used the relation $v_l/u_l=u_{l+1}$;
see the definitions (\ref{eq.2.11}) and (\ref{eq.2.12}).
In a similar way we have $V_2$ as
\begin{eqnarray}
V_2 &=&  E(u_l)
    \sum_{A=1}^{L-3} \ \
    \sum_{B=1}^{L-A-2}
	u_l^A \left ( 1- u_l^{L-1-A-B} \right) \no \\
 &=&	 E(u_l) \sum_{A=1}^{L-3}
   \left[ u_l^A \left( L-2-A \right) + \frac{1}{1-u_l} \left( u_l^{L-1}
-u_l^{A+1} \right) \right]  \no \\
 &=&   E(u_l)^2 \left\{
                 \left( 1+u_l \right) (L-3) u_l^{L-3} \right.\no \\
  & &     \left.         + \left(1-u_l^{L-3} \right)
         \left[  L-2 - \frac{E(u_l) }{u_l} -E(u_l) \right] \right\}.
\label{eq.b.550}
\end{eqnarray}

The calculations of $\phi_2$, $\phi_3$ and $\phi_4$ are done in the same way.
By using eq.\ (\ref{eq.2.18}),
we can write down the function $\phi_2$ in the form
\begin{eqnarray}
\frac{1}{12} \phi_2 (T,L)
&=& \sum_{l=0}^{\infty} \lambda_l^L
      \left\{ f_l \left[   V_3(K,L) + 2 V_4(K,L) +  V_5(K,L) \right]\right. \no
\\
 & & \left. + g_l \left[   V_6(K,L) +   V_5(K,L) \right] \right\}
,\label{eq.b.560}
\end{eqnarray}
where we have
\begin{eqnarray}
V_3(K,L) &\equiv&
    \sum_{C=1}^{L-2} \ \
    \sum_{B=1}^{L-1-C}
      u_l^B v_l^C u_l^{L-B-C}    \no \\
     &=&   E\left( u_{l+1} \right)
            \left[
             ( L-2 )u_l^{L}
              - u_l^2 E\left( u_{l+1} \right)
              \left( u_l^{L-2}-v_l^{L-2}\right)
            \right] , \no \\
            & & \label{eq.b.570}
\end{eqnarray}
\begin{eqnarray}
V_4(K,L) &\equiv&
    \sum_{C=1}^{L-2} \ \
    \sum_{B=1}^{L-1-C}
          v_l^B u_l^C    \no \\
     &=&  E(v_l) \left[ E(u_l) \left(1 - u_l^{L-2} \right)
                    - v_l E\left( \frac{1}{u_{l+1}} \right)
                      \left( v_l^{L-2} - u_l^{L-2} \right)
                 \right], \no \\
                 & & \label{eq.b.580}
\end{eqnarray}
\begin{eqnarray}
V_5(K,L) &\equiv&
    \sum_{C=1}^{L-2} \ \
    \sum_{B=1}^{L-1-C}
          u_l^B u_l^C     \no \\
     &=&  E(u_l) \left[ E(u_l) \left(1 - u_l^{L-2} \right)
                    -(L-2) u_l^{L-1}
                 \right], \label{eq.b.590}
\end{eqnarray}
and
\begin{eqnarray}
V_6(K,L) &\equiv&
    \sum_{C=1}^{L-2} \ \
    \sum_{B=1}^{L-1-C}
      u_l^B     \no \\
     &=&    E(u_l)  \left[ L-2 - E(u_l) \left( 1- u_l^{L-2} \right) \right].
     \label{eq.b.600}
\end{eqnarray}
The function $\phi_3$ is given by
\begin{equation}
\frac{1}{4} \phi_3(T,L)
= 2 \sum_{l=0}^{\infty} \lambda_l^L
	\left\{   f_l \left[   V_7(K,L) +   V_8(K,L) \right]
   +   g_l V_8(K,L) \right\} , \label{eq.b.610}
\end{equation}
where
\begin{equation}
V_7(K,L) \equiv
    \sum_{C=1}^{L-1}
	v_l^C u_l^{L-C}
 = u_l E\left( u_{l+1} \right)
  \left( u_l^{L-1} -v_l^{L-1} \right) , \label{eq.b.620}
\end{equation}
and
\begin{equation}
V_8(K,L) \equiv
   \sum_{C=1}^{L-1}
	u_l^C
 =   E(u_l) \left(1 -u_l^{L-1} \right) . \label{eq.b.630}
\end{equation}
The function $\phi_4$ is written as
\begin{eqnarray}
\frac{1}{3} \phi_4(T,L)
&=& \sum_{l=0}^{\infty} \lambda_l^L
\left\{ 2f_l \left[ (L-1) u_l^L + E(v_l) \left(1-v_l^{L-1} \right) \right]
\right.
\no \\
&&
\left.
+ g_l  \left( 1+u_l^L \right) (L-1) \right\} .
	 \label{eq.b.640}
\end{eqnarray}
We thereby have the expression of the function $\Phi$ in (\ref{eq.b.200}).

We can calculate the function $\Psi$ in (\ref{eq.b.300}) much simpler.
Using eq.\ (\ref{eq.2.17}),
we obtain the explicit form of  $\Psi$ as follows:
\begin{eqnarray}
\frac{3Z_L^2}{L^2} \Psi(T,L) &=& \left\{ \sum_{l=0}^{\infty}(l+1) \lambda_l^L
\left[   1+u_l^L + \sum_{M=1}^{L-1} \left(  u_l^M + u_l^{L-M} \right) \right]
\right\}^2 \no \\
&=& \left\{ \sum_{l=0}^{\infty}(l+1) \lambda_l^L  \left[ 1+u_l^L + 2E(u_l)
\left(1-u_l^{L-1} \right)  \right] \right\}^2 . \no \\
& &  \label{eq.b.650}
\end{eqnarray}

The above expressions (\ref{eq.b.200})-(\ref{eq.b.650}) give a final expression
of the third-order nonlinear susceptibility.
The expressions in the thermodynamic limit are given in ref.\ 3.

Finally, let us explain the calculation of the scaling limit of the
nonlinear susceptibility, namely eq.\ (\ref{eq.2.21}).
The scaling limit is the limit $ K \rightarrow \infty $
with $y=K/L$ fixed.
It is important to notice that the functions $u_l$ and $v_l$ converge to unity
in the scaling limit, except for the terms $u_l^L$ and $v_l^L$, which turn
out to be exponential functions;
see eqs.\ (\ref{eq.2.13})-(\ref{eq.2.16}).

The functions $E(u_l)$ and $E(v_l) $, on the other hand, behave
as follows at low temperatures:
\begin{eqnarray}
E(u_l) &=& \frac{u_l}{1-u_l}
	 \simeq \frac{K}{l+1}
, \label{eq.b.660} \\
E\left( \frac{1}{u_l} \right) &=& \frac{\frac{1}{u_l} }{1-\frac{1}{ u_l} }
	 \simeq - \frac{K}{l+1}
,\label{eq.b.661} \\
E(v_l) &=& \frac{v_l}{1-u_l}
	 \simeq \frac{K}{2l+3}
.\label{eq.b.670}
\end{eqnarray}
These terms partly contribute to the prefactor $J^3/T^6$ of $\chi_3$
in (\ref{eq.1.200}).
Since we factored out $T^{-3}$ from $\chi_3$ in the expression
(\ref{eq.b.100}), we pick out the terms of the order of $J^3/T^3=K^3$ from
the functions $\Phi$ and $\Psi$.
(The factor $1/L$ in (\ref{eq.b.100}) is cancelled out by
the factor $L$ in (\ref{eq.b.400}).)
For example, the function $V_1$ in (\ref{eq.b.540}) is reduced in the scaling
limit to the following form:
\begin{eqnarray}
\label{eq.b.1000}
V_1&\simeq&
\frac{K}{l+1}
\left\{
-\frac{K}{l+2}\frac{K}{y}{\rm e}^{-(l+1)/y}
+\frac{K}{l+1}\frac{K}{2l+3}\left(1-{\rm e}^{-(l+1)/y}\right)
\right.
\nonumber\\
&&
\left.
+\left(\frac{K}{2l+3}-\frac{K}{l+2}\right)\frac{K}{l+2}
\left({\rm e}^{-(2l+3)/y}-{\rm e}^{-(l+1)/y}\right)
\right\}.
\end{eqnarray}
It is notable that the functions $\phi_2$, $\phi_3$ and $\phi_4$
yield terms only of lower orders of $K$ than $K^3$, and hence
vanish in the scaling limit.
Thereby we arrive at  eq.\ (\ref{eq.2.21}).

\pagebreak

\pagebreak
{\bf Figure Captions}

\begin{enumerate}
\item  The spin configuration for the two-point correlation function
 of Heisenberg chain with the periodic boundary condition.
Total number of spin is $L$.
\label{f.1}

\item  The spin configuration for the four-point correlation function
 of Heisenberg chain with the periodic boundary condition.
Total number of spin is $L$.
\label{f.2}



\item
The correction to finite-size scaling of $ \chi_3 T^6 /J^3  $ for
$y= J/(TL)$  =   1.0, 1.2, 1.4, 1.6, 1.8, 2.0
in the case of
the $S=1/2$ ferromagnetic Heisenberg chain
with the periodic boundary condition.
We used  systems with $L$ up to  14.
We assumed that the leading correction is of the form $\sqrt{T/J}$.
Each solid line indicates the extrapolating function
based on the four points nearest to the ordinate.
\label{f.5}

\item
The correction to finite-size scaling of $ \chi_3 T^6 /J^3  $ for
$y= J/(TL) = $ 1.0, 1.2, 1.4, 1.6, 1.8, 2.0
in the case of
the $S=1$ ferromagnetic Heisenberg chain
with the periodic boundary condition.
We used  systems with $L$ up to  10.
We assumed that the leading correction is of the form $\sqrt{T/J}$.
Each solid line indicates the extrapolating function
based on the four points nearest to the ordinate.
\label{f.6}

\item
The finite-size scaling function
$  \tilde{\chi_3}^{\rm peri}(y)$
of the ferromagnetic Heisenberg chain with the periodic boundary condition,
(\ref{eq.2.21}).
The solid curve indicates the scaling function  $ {\tilde \chi_3}^{\rm peri} $
analytically obtained for $S= \infty.$
The numerical data,
which are extrapolated in Figs.\ \ref{f.5} and \ref{f.6},
are indicated by circles for $S=1/2$ and by crosses for $S=1.$
We can obtain the value in the thermodynamic limit,
$  \tilde{\chi_3}^{\rm peri}(y=0)$
for $S=1/2,$ by the Bethe-Ansatz method as in (\ref{eq.1.4}).
\label{f.7}

\item
The correction to finite-size scaling of $ \chi_3 T^6 /J^3  $ for
$y= J/(TL)$  =  0.6, 0.8, 1.0, 1.2, 2.0
in the case of
the $S=1/2$ ferromagnetic Heisenberg chain
with the open boundary condition.
We used systems with $L$ up to  14.
We assumed that the leading correction is of the form $\sqrt{T/J}$.
Each solid line indicates the extrapolating function
based on the four points nearest to the ordinate.
\label{f.8}

\item
The correction to finite-size scaling of$ \chi_3 T^6 /J^3  $ for
$y= J/(TL)$  =  0.6, 0.8, 1.0, 1.2, 2.0
in the case of
the $S=1$ ferromagnetic Heisenberg chain
with the open boundary condition.
We used systems with $L$ up to  10.
We assumed that the leading correction is of the form $\sqrt{T/J}$.
Each solid line indicates the extrapolating function
based on the four points nearest to the ordinate.
\label{f.9}

\item
The finite-size scaling function
$  \tilde{\chi_1}^{\rm open}(y)$
of the ferromagnetic Heisenberg chain with the open boundary condition,
(\ref{tchi3}).
The solid curve indicates the scaling function  $ {\tilde \chi_3}^{\rm open} $
analytically obtained for $S= \infty.$
The numerical data,
which are extrapolated in Figs.\ \ref{f.8} and \ref{f.9},
are indicated by circles for $S=1/2$ and by crosses for $S=1.$
\label{f.10}

\end{enumerate}
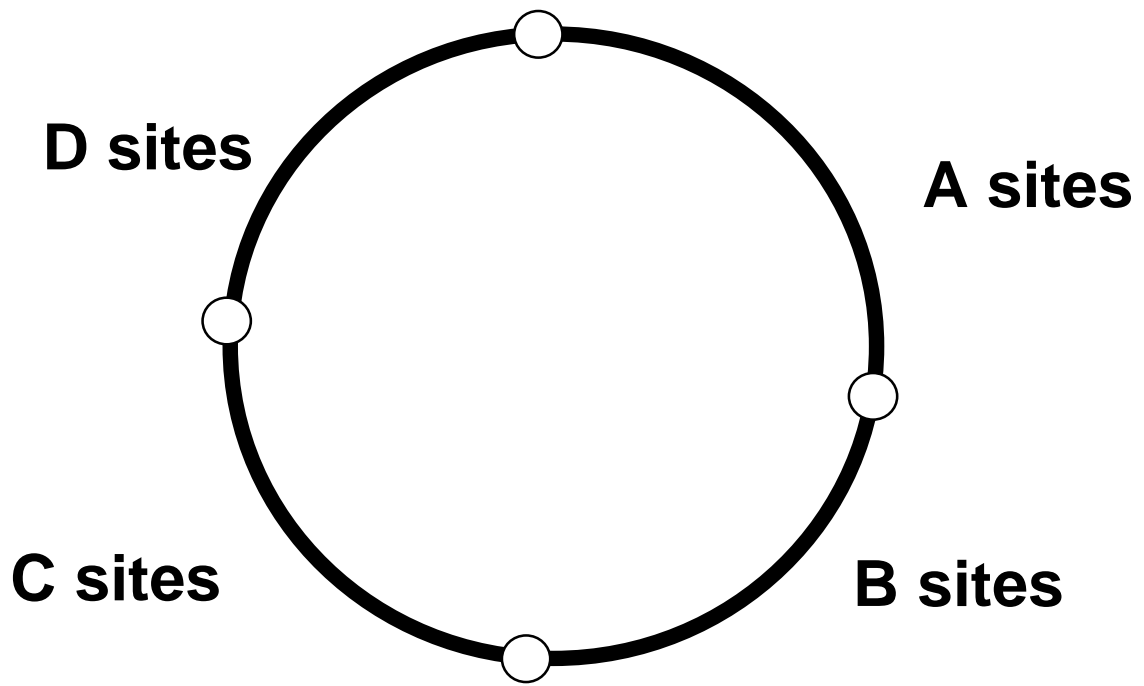
\end{document}